\documentclass[apj]{emulateapj}
\usepackage{times,color}

\def\mbf#1{\mbox{\boldmath ${#1}$}}
\def\Alfven{Alfv\'{e}n~}
\def\eg{\hbox{e.g.}}
\def\ie{\hbox{i.e.}}
\def\etal{\hbox{et~al.}}

\slugcomment{ApJ, in press}

\shorttitle{Fast Waves in Magneto-rotational Winds}
\shortauthors{Suzuki et al.}

\begin{document}

\title{Collisionless Damping of Fast MHD Waves in Magneto-rotational Winds}

\author{T. K. Suzuki$^1$, H. Yan$^2$, A. Lazarian$^2$, \& J. P. Cassinelli$^2$}

\altaffiltext{1}{Department of Physics, Kyoto University, Kitashirakawa, 
Kyoto, 606-8502, Japan; JSPS Research Fellow}
\altaffiltext{2}{Department of Astronomy, University of Wisconsin,
475 N. Charter St., Madison, WI 53706
}

\email{stakeru@scphys.kyoto-u.ac.jp}
\email{yan, lazarian, cassinelli@astro.wisc.edu}

\begin{abstract}

We propose collisionless damping of fast MHD waves as an important mechanism 
for the heating and acceleration of winds from rotating stars. 
Stellar rotation causes magnetic field lines anchored at the surface 
to form a spiral pattern and magneto-rotational winds can be driven.
If the structure is a magnetically dominated, fast MHD waves
generated at the surface can propagate almost radially outward and
cross the field lines.  The propagating waves undergo collisionless damping 
owing to interactions with particles surfing on magnetic mirrors that are 
formed by the waves themselves.  The damping is especially 
effective where the angle between the wave propagation and the field
lines becomes moderately large ($\sim 20$ to $80^{\circ}$). The angle tends
naturally to increase into this range because the field in magneto-rotational winds
develops an increasingly large azimuthal component.
The dissipation of the wave energy produces  heating and acceleration of the outflow.
We show using specified wind structures that this damping process can be important in 
both solar-type stars and massive stars that have moderately large rotation 
rates. 
This mechanism can play a role in coronae of young solar-type stars which 
are rapidly rotating and show X-ray luminosities much larger than the sun.
The mechanism could also be important for producing the
extended X-ray emitting regions inferred to  exist in massive stars of
spectral type middle B and later.
\end{abstract}

\section{Introduction}
\label{sc:intro}

A wide range of stellar classes are thought to have hot plasmas with
temperatures greater than $10^6$ K in their atmosphere. Main sequence stars
with low to intermediate mass, including the sun, are known to possess hot
coronae, and stellar winds emanate from coronae by thermal expansion most
likely enhanced by magnetic wave energy deposition. Magnetic fields
generated by dynamo mechanisms in these stars are likely brought to the
surface by magnetic buoyancy in the convection zones \citep{pak66}, and the
fields play a major role in the structure of the outer atmosphere. In a
static atmosphere, the gas pressure, $p$, falls exponentially with a scale
height $H$, but the magnetic field, $B$, generally decreases more slowly,
as a power-law with radius $r$, to satisfy conservation of magnetic flux. 
As a result, the magnetic pressure, $B^2/8\pi$, can eventually dominate 
thermal pressure from a certain radius outward, even though the thermal 
pressure dominates in regions nearer the photosphere. Thus in the outer 
regions the dissipation of the magnetic energy can greatly influence the 
energetics and dynamics of the atmosphere and wind, producing in some way, the 
heating of coronae and the wave acceleration of stellar winds. We intend 
to explore a wave damping process that can produce the heating and 
acceleration.

Stellar rotation is also closely related to the coronal activities.  
The observed X-ray luminosity of low to intermediate mass stars 
increases monotonically with decreasing rotation period, $P$ 
for $P \gtrsim$ several days, and then is saturated for smaller 
values of $P$ \citep{piz03}. 
It is likely that the stronger differential rotation in the interior 
leads to the generation of a surface magnetic field 
that is stronger, and a more active corona is formed. 
However, even qualitative arguments are still lacking as to how the 
heating and acceleration by the wave damping actually occurs in the
atmospheres of rotating stars.    

It is now becoming more evident that the heating and driving effects of
magnetic fields is not confined only to cool stars, but can be important in
the more massive early type stars as well. For many years it had been
believed that dynamo generated magnetic fields would not exist on hot stars,
and that the winds from these objects were produced only as a result of
radiation pressure gradients. Unlike the cooler stars, the hot massive stars
do not have an outer convection zone, so the mechanism for the rise of
magnetic fields and the source of mechanical energy for coronal heating are
not present. However, early UV spectra of the stars showed lines from
anomalously high ionization stages \citep{lm76}. This superionization
could be explained as resulting from X-ray ionization by the Auger Effect 
\citep{co79}. It was soon realized from
satellite observations that essentially all O and B stars are X-ray sources 
(Seward et al. 1979; Bergh\"{o}fer et al. 1997), 
but the nature of the source of the X-rays 
remained a mystery. \citet{cs83} provided arguments that perhaps there are both
shocks in the winds (as proposed by Lucy 1982), as well as hot coronal regions 
at the base of the winds. This idea that there are two contributors to the 
X-ray emission has persisted (Waldron and Cassinelli, 2000). Also, in the case of 
B near main sequence stars,  Cohen \etal (1997) found that the X-ray emission 
measure was comparable-to or larger than that of
a theoretically predicted wind, and hence the coronal component of the X-ray
emission could be dominant for these stars. As for the magnetic fields in
the outer atmospheres and winds, some direct measurements have been made in
near main sequence B stars with fields of about 300 Gauss (Donati \etal\ 2001). 
Charbonneau and MacGregor (2001) have shown that fields can be generated by a dynamo
operating at the interface between the convective core and the radiative
envelope in early type stars. MacGregor and Cassinelli (2003) have shown
that such fields can buoyantly rise through the radiative zone to the
surface of hot stars. In addition there is a wide variety of observational
evidence that hot stars have magnetic fields (Henrichs 2001). 

\begin{figure}
\figurenum{1} 
\epsscale{.8} 
\plotone{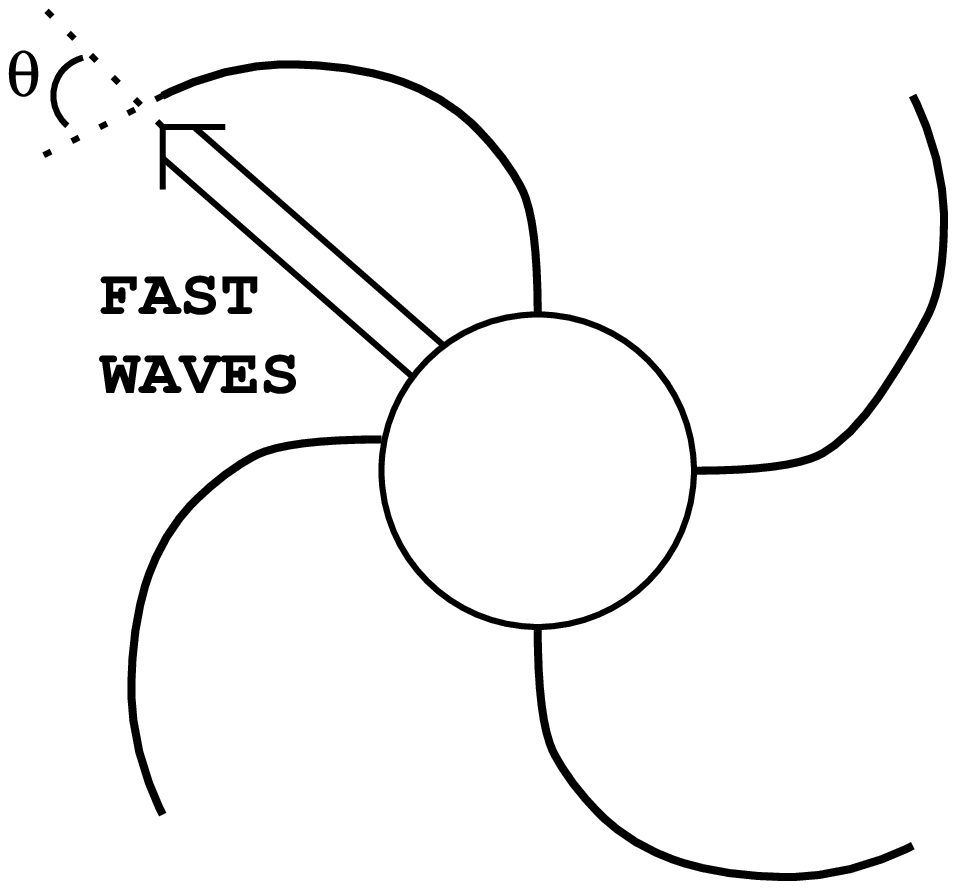} 
\caption{Schematic picture of fast MHD waves propagating outward
in magneto-rotational wind, illustrating the angle $\theta$ between the 
propagation direction and the field at one point in the outer atmosphere.}
\label{fig:scpc1}
\end{figure}

Thus, there
appears to be a greater similarity between hot and cool stars than had been
envisioned, and it is both timely and useful to consider the processes
by which magnetic fields can heat and accelerate winds across the HR diagram.
Here the focus is on structures in which the field in the wind
is configured in the form of a spiral, as would naturally arise 
from the rotation of a star with a magnetic field rooted in its
surface (fig. \ref{fig:scpc1}) \citep{wd67,bm76}.

Magnetohydrodynamic (MHD) waves are widely regarded as promising
mechanisms that play a role in the heating and acceleration of the outer
atmospheric plasmas. Surface turbulence and transient activities (e.g.
Sturrock 1999) excite MHD waves at the surface of the sun. It is natural to
expect that similar excitation phenomena occur on other stars that have
surface magnetic fields. Outwardly propagating MHD waves carry Poynting flux
energy, and any dissipation of this flux directly leads to conversion from
magnetic energy to thermal and kinetic energy in upper atmosphere. There are
numerous studies of wave heating in the solar corona / wind (e.g. Belcher
1971) and in stellar coronae / winds (e.g. Hartmann \& MacGregor 1980).
MHD waves are discussed in regards to pulsar winds as well \citep{lyu03}.

A flow of wave-like motions created by turbulence in a 
localized source can be decomposed into three types of MHD waves: 
fast, \Alfven, and
slow modes (see Cho \& Lazarian 2003). In a magnetically dominated plasma,
the slow mode essentially corresponds to an acoustic wave, hence, it
probably does not contribute the dominant amount of heating to the plasma in
the absence of other strong driving mechanisms (as discussed in
\S\ref{sc:otmd}). The fast and \Alfven modes, however, can play a dominant
role since their wave energies can easily exceed the thermal energy. These
two modes propagate quite differently. The \Alfven waves propagate as
transverse waves and only along the field lines, while fast waves can 
propagate almost isotropically. The properties of fast mode wave 
do depend on the direction of the wave relative to the magnetic field line
and thus can change with radial distance from the star.
The component of the fast mode wave which is traveling along the field line
is transverse, as is the case for the \Alfven mode, but the fast wave 
becomes increasingly compressive as the angle, $\theta$, between the wave
propagation vector and the field line increases.

Some of the properties of the wave propagation are affected by the rotation
of the star 
because the rotation causes the magnetic field lines that are rooted on the 
star to form a spiral pattern as described above.  
In the solar wind, the field is 
usually pictured as developing
a so-called Parker spiral \citep{pak63}. This spiral arises in spite of the
relatively slow rotation of the Sun. More tightly wrapped spiral patterns
should form in the winds of stars that rotate faster, such as younger
solar-mass stars and massive stars. Thus, it is important and interesting to
study outward propagation of the fast MHD waves in magneto-rotational winds
of rapidly rotating stars in general.

Let us consider outgoing fast waves emergent from the stellar 
surface.  
The fast waves will travel almost radially without being
refracted provided that the magnetic energy in the plasma dominates the
thermal energy. However, as the wave travels away from the surface, the
angle of propagation, $\theta$, increases monotonically (fig. \ref{fig:scpc1}) 
and the character of the wave changes accordingly. 
The way in which the dissipation of the waves
is influenced by this variation of $\theta$ is described in (\S
\ref{sc:dampfs}), where we also discuss how the stellar rotation plays a
dominant role in controlling the wave dissipation and subsequent heating of
the stellar atmosphere (\S \ref{sc:rslt}).

An important aspect in the discussion of stellar plasmas is that they
are generally collisionless except in regions very near the surface. 
This is the case because the mean free path for a Coulomb collision, 
\begin{equation}
\label{eq:clsth}
l_{\rm mfp} = 
9.38\times 10^{7}{\rm cm}\frac{(T/10^6{\rm K})^2}{(n/10^8{\rm cm}^{-3})}, 
\end{equation}
is larger than wave length. Thus it is essential for us to consider
collisionless processes in the wave dissipation. 
The fast mode waves suffer collisionless damping 
\citep{gin61,bar66}, which is discussed in the context of the heating 
of the solar wind plasma by \citet{bar68,bar69}. 
Interestingly, the damping rate sensitively depends on $\theta$
(\S\ref{sc:dampfs}). To date, however, this damping process has not been
applied to magnetically dominated winds 
from rapidly rotating stars. 
It is the goal of this paper to
investigate the importance of collisionless damping for both the heating and
acceleration of stellar coronae in which the field has the configuration of
a magneto-rotational wind.

\section{Damping of Fast MHD Waves}
\label{sc:dampfs}

In this section, we summarize the physical process of collisionless
damping of the fast MHD waves, and we introduce the damping rate.
It is useful to compare the collisionless damping process  with the
one that has often been considered in the literature \citep{hol82,suz04};
the damping of waves through the formation of a fast shock train by the 
steepening of the wave fronts.  We find circumstances in coronae under 
which the collisionless damping can become more important than the shock 
steepening process.

\begin{figure}
\figurenum{2} 
\epsscale{1.} 
\plotone{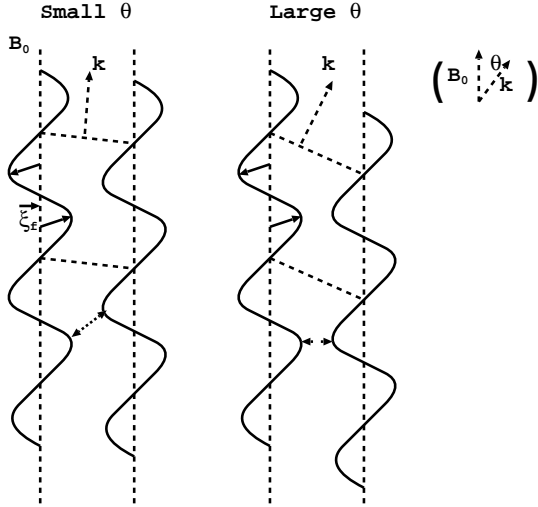} 
\caption{
Propagation of fast MHD waves for two different angles, $\theta$. 
In a low $\beta$ plasma, the displacement vector, $\vec{\xi}_f$, of the fast 
mode is  almost perpendicular to underlying magnetic field, $B_0$ \citep{clv03}. 
This illustrates that the distance between the two oscillating field lines 
(solid lines) changes more in larger $\theta$ case 
than in the smaller $\theta$ case. This is the case even though, as illustrated,
the absolute displacements, $|\vec{\xi}_f|$, 
are the same. Particularly, the closest distance between 
the two field lines (indicated by dotted both-side-arrows) is smaller 
in the large $\theta$ case,   
leading to a more reflective mirror. 
Hence, a larger number of particles
are locked into the surfing on the mirrors. This in turn leads to a 
higher rate of wave dissipation by particle-field interactions.}
\label{fig:fstmir}
\end{figure}

As discussed in Ginzburg (1961), the nature of collisionless damping of 
a wave can be considered analogous to the creation of radiation by charged 
particles  in magnetic field. The charged particles can emit electromagnetic 
waves both by their acceleration (which produces cyclotron radiation) and by the 
Cherenkov effect.  By the inverse of these two processes charged particles can 
absorb the radiation. Similarly,  plasmas
can absorb energy from plasma waves by a gyroresonance process or by a 
process analogous to the Cherenkov effect. It is the latter that is of interest
to us here. 
It leads to a dissipation of the plasma waves 
by a wave-particle interaction, which 
results in  heating and acceleration of the particles that compose the plasma. 

The resonance of a wave with the 
gyro-resonance frequency of ambient thermal ions causes a damping of 
those wave modes that have a frequency close to the ion-cyclotron frequency 
(e.g. Leamon et al.1998).
Thermal particles can also be accelerated by an oscillating parallel 
electric field, as is the case for Landau damping of a wave. 
A similar mechanism 
called transit time damping (TTD) 
operates if the electric field is replaced by moving magnetic mirrors.  
\citet{bar66} 
introduced this idea to explain the heating 
of the solar wind, and \citep{yl04} applied to studies on cosmic ray 
scattering. 
In this paper, we consider the TTD of fast MHD waves 
in rapidly rotating stars. 
We can reasanably assume a magnetically dominated medium, commonly
called a low $\beta $ plasma. As usual, $\beta $ is the
the ratio of the gas pressure  $p_g = \rho a^2 $, to the magnetic pressure, 
$p_b = B^2 / 8\pi$, 
\begin{equation}
\label{eq:plsmbt}
\beta \equiv \frac{p_{g}}{p_{b}} = \frac{8\pi \rho a^2}{B^2} ,  
\end{equation}
where $B$ is the magnetic field strength, $\rho$ is the ambient gas density, 
and $a$ is the isothermal sound speed.   

The TTD is caused by interactions between magnetic 
mirrors associated with the fast waves (Figure \ref{fig:fstmir}) and particles 
that have a parallel  velocity, $v_{\parallel}$, similar to the pattern 
speed of the  mirrors, that is,
\begin{equation}
v_{\parallel} \simeq  \omega / k_{\parallel} \; .
\end{equation}
This is the Cherenkov condition, and $\omega$ and 
$k_{\parallel}$ are the wave 
frequency and the wave number parallel to the magnetic field lines. 
Particles which satisfy this condition surf on the mirrors 
and thus interact with the waves. 
The damping rate $\gamma_{d}$ 
of fast waves which propagate with an angle $\theta (\simeq 1$ radian) 
relative to the field lines in a plasma of a given $\beta(\ll 1)$, and for 
the Maxwell-Boltzmann particle distribution is \citep{gin61} 
$$
\hspace{-4cm}
\gamma_{d} = \frac{\sqrt{\pi\beta}}{4}\omega\frac{\sin^{2}\theta}{\cos\theta}
$$
\begin{eqnarray}
\label{eq:damp}
\times\left[\sqrt{\frac{m_{e}}{m_H}}\exp(-\frac{m_{e}}
{m_H\beta\cos^{2}\theta}) 
 + 5\exp(-\frac{1}{\beta\cos^{2}\theta})\right],
\end{eqnarray}
where $m_H$ and $m_{e}$ are the hydrogen and electron masses, 
and this expression is strictly true only for linear waves in 
which the velocity amplitude is smaller than the phase speed. 
Note that $\theta$ changes owing to the turbulent fluctuations in actual
situations \citep{lv99,yl04}. However, since we are considering small
relative amplitude, $\delta B/B$, and are primarily interested in the
average properties, we choose here to treat $\theta$ as the average angle
between the underlying magnetic field and the wave propagation direction. 

\begin{figure}
\figurenum{3} 
\epsscale{.9} 
\plotone{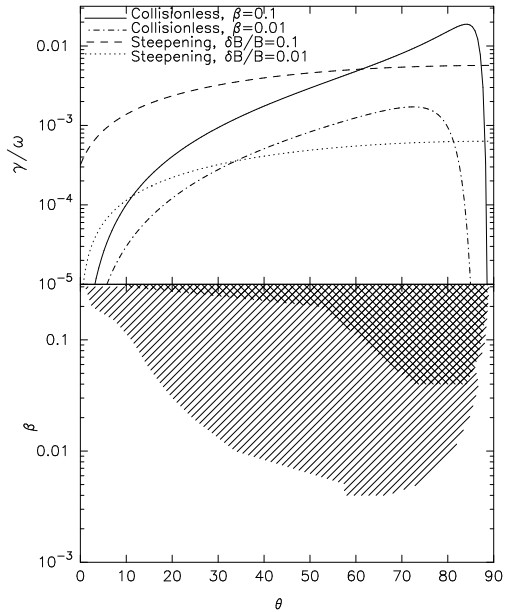} 
\caption{{\it upper}: Comparison of damping rate of the collisionless process 
with the steepening as a function of $\theta$. 
{\it lower} : Regions in $\theta - \beta$  where the collisionless process 
dominates the steepening for $dB/B = 0.1$ (cross hatched) and $dB/B = 0.01$ 
(single hatched).}
\label{fig:damp}
\end{figure}

The Upper panel of fig. \ref{fig:damp} presents $\gamma_d$ divided by
$\omega$ as a function of the angle $\theta$ for two values of $\beta$;
$\beta=0.1$ and 0.01. This collisionless damping is shown in comparison with
the damping rate associated with a shock train that is usually used for such
discussions. (see the Appendix for a discussion of the the damping of waves
by the shocks). The rate of the collisionless damping is independent of wave
amplitude, $\delta B$, because it is a linear process, while it is larger 
in plasma which have a $\beta$ value not too far below unity.
This is because the parallel speeds of the surfing particles needs to be
comparable to fast mode speed for the resonance to occur. If $\beta$ is too
low (\ie the sound speed is too low), only an insignificant fraction of the
thermal particles will have the large
values required for the condition given in eq.(3). 

Collisionless damping 
is also quite sensitive to the angle $\theta$. The damping rate increases 
more rapidly than the steepening 
in the range, $\theta \lesssim 80^{\circ}$,  
but then it suddenly decreases as $\theta$ comes closer to $90^{\circ}$.    
This dependence on $\theta$ 
can be understood by considering two competing factors. In general, the
damping increases with $\theta$ because a magnetic mirror tends to be formed
in a more efficient way, as is illustrated in Figure \ref{fig:fstmir}. 
A small separation of the field lines is an indication that a magnetic mirror 
has formed. 
The closest distance between the field lines 
become smaller for larger $\theta$ so that the mirrors become more reflective 
even though the amplitudes of the fast waves are the same. 
Thus, more particles are interacting with the wave to cause the dissipation. 
However, when
$\theta$ approaches to $90^{\circ}$, $k_{\parallel}$ becomes small, and the
resonance condition (eq. 3) requires a large $v_{\parallel}$. Hence, most
thermal particles will no longer be in resonance with the wave, so the
damping rate decreases sharply in the $\theta$ near $90^{\circ}$ regime.
\footnote{The former effect appears in the $\frac{\sin^2 \theta}
{\cos\theta}$ term
in eq.(\ref{eq:damp}) which increases with $\theta$, while the latter effect 
appears in the exponential terms in eq.(\ref{eq:damp}) which both 
approach 0 as $\theta \rightarrow 90^{\rm o}$. }

On the other hand, the damping caused by the shocks 
does depend on $\delta B$ because it is
a nonlinear process, and  is almost independent of the plasma
$\beta$ value 
(see Appendix eq.(\ref{eq:dmpscl})). So, 
we adopt $\beta = 0.01$ for the two cases of the shock damping. 
The damping rate by the shock steepening also increases with 
$\theta$ as in the collisionless process, 
because fast waves contain a longitudinal, {\it i.e.} compressional, 
component as $\theta$ increases, and they become more dissipative. However, 
the variation is more
weakly dependent on $\theta$ than is the case of collisionless damping.

The conditions under which the collisionless process exceeds
the shock steepening process is further illustrated in the lower panel of 
fig. \ref{fig:damp} 
which shows on a $\theta$ versus $\beta$ plane for two cases, $\delta B/B=0.1$ 
(cross hatched) and $\delta B/B=0.01$ (single hatched). 
Note that the collisionless process tends to become more important
for larger values of $\theta$, except when $\theta$ is near $90^\circ$ and/or
when the $\beta$ value is too small. 
As is shown, a fast wave with amplitude $\delta B/B = 0.01$ will dissipate
predominantly by collisionless damping over a wide range of $\theta$. Even
for the case in which $\delta B/B = 0.1$, the collisionless damping is seen
to be more important for $\beta \simeq 0.1$ in an angular region of
$60^{\circ} \lesssim \theta < 90^{\circ}$. These results show that there are
interesting physical conditions under which collisionless damping should,
from now on, be taken to be the primary process for wave dissipation.

These results have shown that the collisionless damping can be
comparable-to, or more important than, shock steepening for the damping of fast
waves. In fact, collisionless damping could dominate even more-so. This
is because shock steepening could be smaller than we have 
assumed, for the following reasons. First, when a sinusoidal wave is excited 
initially, it takes some time for it to steepen to form the shocks before the shock
dissipation process can begin.  Whereas the collisionless damping works
immediately, even for the case of sinusoidal waves. Second, the
collisionless effects can themselves modify the shock dissipation.  In
general astrophysical circumstances, each shock in a shock train is
considered collisionless, so the plasma is not thermalized by Coulomb
collisions, but rather the plasma is randomized because of fluctuations in
the magnetic field. 
In such situations, particles with energy larger than average 
can move across the front more or less transparently and the conversion from 
kinetic energy to heat would be less effective, and hence also
the shock dissipation would be less.
We conclude from
this section that our results indicate that it is important to account for
collisionless damping of the fast modes for realistic and interesting
astrophysical situations.

\section{Model for Magneto-rotational Winds} 

Let us consider the outward propagation of the fast MHD waves for
the case of  magneto-rotational stellar winds.  
Figure \ref{fig:scpc1} depicts propagation of the fast MHD waves in 
the rotational winds seen from the pole. Here we adopt a simple 
split monopole configuration for the magnetic fields and 
assume that the magnetic axis  coincides with the  stellar spin axis. 
Spiral magnetic fields are formed in the equatorial
region by the stellar rotation  \citep{wd67,bm76}. 
We can also reasonably assume that the ideal MHD condition holds in the 
stellar atmosphere.  
These are common assumptions for the equatorial region of 
a magneto-rotational wind. 

The plasma itself cannot move across the field lines. 
However, the fast MHD waves can propagate almost 
isotropically in a low $\beta$ plasma. In particular, they can propagate  
radially,  crossing the field lines as illustrated in the figure. Also
shown is the increase in the angle $\theta$ as the wave travels away 
from the surface.     
Where the angle $\theta$ is large, the fast waves suffer 
collisionless damping, as we discussed in the previous section 
as long as the wave length, $\lambda$, is short enough to satisfy 
collisionless condition, 
\begin{equation}
\label{eq:clles}
\lambda < l_{\rm mfp}. 
\end{equation}
For example, if we consider a fast mode wave with period of $1$s, with a phase
speed is $1000$ km/s traveling in plasma with a temperature of $2\times
10^6$ K, the collisionless condition that must be satisfied for the particle
density is $n\lesssim 10^8$ cm$^{-3}$. In the case of the sun, this density
condition holds even in the inner corona. As a result, the wave dissipation
process directly leads to heating and acceleration of wind plasma and may
have an effect on the wind structure.

In this section, we present a model to explore the collisionless damping of 
the fast mode for the case of magneto-rotational winds. 
As we are just beginning to develop an understanding of the basic
collisionless damping process, it is too soon to be aiming for a complete
and self consistent picture. For example, there are other processes
such as shock dissipation argued in the previous section that may still be 
important.
We choose to estimate the energy and momentum that is transfered by the 
fast wave dissipation in the case of a pre-specified background wind.
Then we can discuss the likely influence on the wind energetics and 
dynamics.

\subsection{Approximated Wind Structure}
\label{sc:apwdst}

Here we describe a practical method for constructing approximate wind 
structures in the equatorial plane of the magneto-rotational winds,
as discussed in Lamers and Cassinelli (1999).
We assume that the wind is  isothermal at a  temperature $T$,
and we assume that all the physical quantities depend spatially only on  
the radial distance, $r$ in the equatorial plane. 
For the split monopole magnetic field, the conservation of magnetic 
flux, $\mbf{\nabla \cdot B}=0$, fixes  radial component of the magnetic field, 
\begin{equation}
B_r = \left(\frac{R_\star}{r}\right)^2 B_{r,0},  
\end{equation}
where $R_{\star}$ is stellar radius and $B_{r,0}$ is radial field strength
at the surface where $r=R_\star$. It is empirically known that radial
velocity distribution of stellar winds can be represented by the ``beta
velocity law'' (here we use the power $\eta$ to avoid confusion with the 
plasma $\beta$ ratio that we have used throughout), thus
\begin{equation}
\label{eq:vrstr}
v_r = v_\infty \left(1-\frac{R_\star}{r}\right)^{\eta}, 
\end{equation}
where $\eta \simeq 0.5$--3. 
If the rotation is slow, terminal velocity, $v_\infty$, is of order the 
escape velocity, $v_{\rm esc,0} \simeq \sqrt{\frac{2 G M_\star}{R_\star}}$, 
where $M_{\star}$ is stellar mass.

In rapid rotating stars, $v_{\rm esc,0}$ may not a good indicator of 
the terminal velocity, as $v_\infty > v_{\rm esc,0}$ because the wind
is significantly affected by acceleration from the  magneto-rotational forces.
For these cases, Lamers and Cassinelli (1999) show that
$v_\infty$ is better estimated by the 
Michel velocity \citep{mc69}, which is  defined as  
\begin{equation}
v_{\rm M} = \left(\frac{r^4 {B_r}^2 \Omega^2}{\dot{M_{\rm E}}}\right)^{1/3} 
= \left(\frac{R_{\star}^4 {B_{r,0}}^2 \Omega^2}{\dot{M_{\rm E}}}\right)^{1/3}, 
\end{equation}
where $\Omega$ is angular speed of stellar rotation 
and $\dot{M_{\rm E}}$ is equatorial mass flux rate multiplied by $4\pi
r^2 $ and we use, 
\begin{equation}
\dot{M_{\rm E}} = 4\pi \rho v_r r^2 (= {\rm const.})
\end{equation}
In this paper we adopt  
\begin{equation}
\label{eq:tmvl}
v_\infty = \max(v_{\rm esc,0},v_{\rm M}).  
\end{equation}
It should be noted that $v_{\infty}$ can be estimated by the surface values 
since both $v_{\rm esc,0}$ and $v_{\rm M}$ are determined at $r=R_{\star}$. 
From here on, we follow convention and call cases with $v_{\rm esc,0} > v_{\rm M}$,  
$v_{\rm esc,0}\simeq v_{\rm M}$, and  $v_{\rm esc,0} < v_{\rm M}$, the ``slow'', 
``moderate'', and ``fast'' magneto-rotators, respectively.

We can determine azimuthal field component, $B_\phi$, for given 
$\Omega$ by the induction equation as
\begin{equation}
\frac{B_\phi}{B_r} = \frac{v_\phi - r\Omega}{v_r} .
\end{equation}
Azimuthal velocity, $v_{\phi}$, is derived from conservation of 
angular momentum as 
\begin{equation}
v_{\phi} = r \Omega \frac{\frac{v_r^2 {\cal L}}{r^2 \Omega} -v_{{\rm A},r}^2}
{v_r^2 -v_{{\rm A},r}^2} = r \Omega \frac{\frac{r_{\rm A}^2}{r^2}v_r^2 - 
v_{{\rm A},r}^2}{v_r^2 -v_{{\rm A},r}^2},  
\end{equation}
where $\cal L$ is specific angular momentum, $v_{{\rm A},r} = \frac{B_r}
{\sqrt{4\pi \rho}}$ is radial \Alfven speed, and $r_{\rm A}$ is \Alfven
point at which $v_r = v_{{\rm A},r}$. Thus, we have set the structures of
the magneto-rotational winds for given $M_{\star}$, $R_{\star}$,
$\dot{M}_{\rm E}$, $B_{\rm r,0}$ and $S_0$, where $S_0$ is spin normalized
by Kepler velocity, $\Omega_{\rm K}=\sqrt{\frac{G M_\star}{R_\star^3}}$:
\begin{equation}
S_0 \equiv \frac{\Omega}{\Omega_{\rm K}}
\end{equation}

\subsection{Wave Propagation}

We introduce formulation describing propagation of the fast MHD waves 
with dissipation in the magneto-rotational winds under the WKB approximation. 
We only consider magnetically dominated {\it i.e.}, 
low $\beta$ plasmas. 
Accurately speaking, refraction affects the wave propagation; the fast wave 
propagates toward a region with smaller phase speed, $v_{\rm f}$, defined 
as\footnote{Although adiabatic sound speed 
should appear here, we use isothermal sound speed, $a$, as we assume the 
isothermal atmosphere.}
$$
v_{\rm f}= \frac{B}{\sqrt{8\pi\rho}} \left[ 1 + \frac{4\pi\rho a^2}{B^2} 
\right.
$$
\begin{equation}
\label{eq:fstwsp}
\left.
+ \sqrt{(1 + \frac{4\pi \rho a^2}{B^2})^2 - \frac{4B_r^2}{B^2}
\frac{4\pi\rho a^2}{B^2}}\right]^{1/2}, 
\end{equation}
which corresponds to a direction away from the field line, and where
$B=\sqrt{B_r^2 + B_\phi^2}$.
However, in the case of a low $\beta$ plasma we can reasonably assume that 
the fast waves propagate radially without suffering the refraction, 
because $v_{\rm f}\simeq \frac{B}{\sqrt{4\pi\rho}}$ is almost independent 
of $\theta$.

The energy flux of MHD waves in fluids moving with velocity, $\mbf{v}$ is 
written as \citep{jaq77}
\begin{equation}
\mbf{F}_{\rm w} = \mbf{V}_{\rm g} \cal{E}_{\rm w} + \mbf{v \cdot} 
{{\it p}_{\rm w}}
\end{equation}
where ${\cal E}_{\rm w}$ is wave energy density, $\mbf{V}_{\rm g}$ is 
group velocity, and $p_{\rm w}$ is wave pressure.  
For a wave amplitude, $\delta v$, the energy density is ${\cal E_{\rm w}} 
= \frac{1}{2} \rho \delta v^2$.  
The energy flux, $\mbf{F_{\rm W}}$, changes according to work done by $p_{\rm w}$ 
and damping energy loss rate, $-Q_{\rm w}$, so that in  steady state, 
\begin{equation}
\label{eq:wveng}
\mbf{\nabla \cdot F_{\rm w}} - \mbf{v \cdot }(\mbf{\nabla} p_{\rm w}) 
= -\rho Q_{\rm w}.    
\end{equation}
Note that $Q_{\rm w}$ works as a heating term for the surrounding plasma.  
We are focusing on fast MHD waves in the low $\beta$ plasma. 
Thus the radial components of the pressure gradient, 
$\mbf{\nabla} p_{\rm fw}$, and the energy flux, $\mbf{F_{\rm fw}}$, 
can be expressed 
as 
\citep{jaq77}
\begin{equation}
\label{eq:wvprgr}
(\mbf{\nabla \cdot} p_{\rm fw})_r = (\frac{1}{2} + \sin^2 \theta) 
\frac{d {\cal E}_{\rm fw}}{dr} + \frac{3 {\cal E}_{\rm fw}}{r} 
\sin^2 \theta
\end{equation}
and
\begin{equation}
\label{eq:wvenfl}
F_{{\rm fw},r} = {\cal E}_{\rm fw} \left[(\frac{3}{2} + 
\sin^2 \theta)v_r + v_{\rm f}\right].  
\end{equation}
The term $\gamma_{\rm d}$ in eq. (\ref{eq:damp}) is defined as the damping
rate with respect to normalized amplitude, 
$\delta w \equiv \delta v/ v_{\rm f}$. 
Substitutions of eqs.(\ref{eq:wvprgr})
and (\ref{eq:wvenfl}) into eq. (\ref{eq:wveng}) gives an expression for
the variation $\delta w$ in the stellar winds.

\hspace{-7cm}
$$\frac{d}{dr}(\delta w) = - \delta w \left[\frac{\gamma_{\rm d}}
{v_r + v_{\rm f}} + \frac{1}{2\rho} \frac{d\rho}{dr} + \frac{1}{v_{\rm f}} 
\frac{d v_f}{dr} \right.$$
$$
+ \frac{1}{2 (v_r + v_{\rm f})} \left\{ \frac{(3 - \sin^2 \theta)v_r 
+ 2 v_{\rm f} }{r} \right.
$$
\begin{equation}
\label{eq:vrdv}
\left. \left.
+  \frac{d}{dr}\left\{ 
(\frac{3}{2} + \sin^2 \theta)  v_r + v_{\rm f}
\right\} \right\} \right], 
\end{equation} 
The first term on the right hand side denotes the collisionless damping. 
The second term indicates the amplification that occurs in a 
stratified atmosphere in which the density $\rho$ is decreasing. 
The third term appears because $v_{\rm f}$ is used for the normalization 
of $\delta w$. 
The fourth term is due to geometrical expansion of 
flow tubes and the variation of $v_{\rm f}$ 

For a given wave frequency, $\omega$, the damping rate $\gamma_{\rm d} 
\left( = \gamma_{\rm d} (\beta,\theta,\omega) \right)$ is  
determined by the plasma $\beta$ (from eq. \ref{eq:plsmbt}) and the angle $\theta$
between the propagation direction and the field line, for the assumed radial
propagation, that is defined by
\begin{equation}
\theta = \tan^{-1}\left(\frac{B_\phi}{B_r}\right) .
\end{equation}
Thus we see that the damping rate, $\gamma_{\rm d}$, is fully determined by the 
background wind properties. 

The initial conditions of the outgoing fast mode waves,
at the surface of the star, provide the amplitude, $\delta w_0 \equiv 
\delta v_0/v_{\rm f,0}$, and the period, $\tau(=2\pi/\omega)$.
In the following calculations, we consider the results for the specific
initial amplitude $\delta w_0 =0.1$. In a region close to the surface, the plasma is
generally collisional, i.e. $\lambda (= v_{\rm f} \tau) < l_{\rm mfp}$. In
this region, eq. (\ref{eq:vrdv}), for collisionless damping, is not
applicable, so we assume $\gamma_{\rm d} = 0$ when solving
eq. (\ref{eq:vrdv}). However, even without the damping, $\delta w$ decreases 
because of the rapid increase of $v_{\rm f}$, (see \S \ref{sc:rslt}). As a result,
all other damping processes should not be effective either, so our treatment has
acceptably isolated the collisionless damping case.
Once the collisionless condition, eq. (\ref{eq:clles}), is satisfied at 
$r=r_{\rm cl}$ and beyond, we can start the outward integration of 
eq. (\ref{eq:vrdv})  to determine $\delta w$.   
Using this, we can estimate heating, $Q_{\rm w}$, and find the acceleration, 
$-\frac{dp_{\rm w}}{dr}$, that is caused by the dissipation of the fast 
MHD waves.

\section{Results}
\label{sc:rslt}

For our model of collisionless dissipation of fast MHD waves in a
magneto-rotational wind, there are two important requirements.
First, a sufficiently large rotation rate is required to create 
a sufficiently large $\theta$ for the wave dissipation to occur.
Second, stellar winds with moderately low plasma $\beta$ values, 
(\ie\ $\beta \sim 0.01 - 0.1$) are favored, because
the fast waves, which are essentially magnetic waves, can be
dominant only for low $\beta$ plasmas, although the dissipation
is suppressed if the $\beta$ value is too small.
(eq. \ref{eq:damp}).

\subsection{Solar-type Stars}

Moderately low $\beta$ winds are found in the coronal winds of
low to intermediate mass stars. 
Although the Sun is certainly one of the candidates in the sense that it
is a moderate to late type star, it has a rotation rate
that is too slow to produce the large $\theta$ assumed in our 
discussion.  
Observations of F-K (mid- to late-type) stars in open clusters 
\citep{bar03} show that stars with age $\lesssim 100$ Myr have as large spin
\footnote{$S_0=0.005$ for the sun.} as $S_0 \simeq 0.5$.  
Thus these are good classes of stars for us to be considering here.

\begin{figure}
\figurenum{4} 
\epsscale{1.} 
\plotone{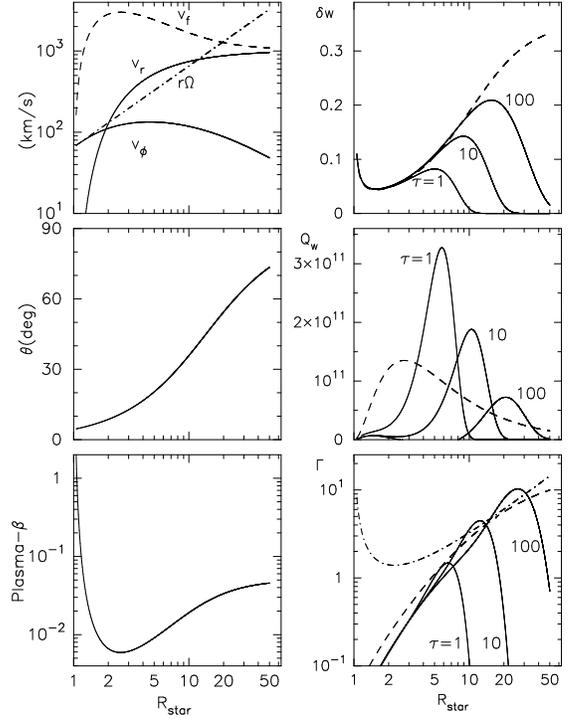} 
\caption{Wave dissipation in a solar-type star with $S_0 =0.15$. All the 
physical quantities are as a function of stellar radius. 
{\it Upper left} : velocity structure. {\it Middle left} : $\theta$. 
{\it Lower left} : plasma $\beta$ value. 
{\it Upper right} : Normalized amplitude of the fast waves. Solid lines 
are results with dissipation for three cases of wave period, $\tau$. 
Dashed line is a result without dissipation. {\it Middle right} : 
Comparison of heating, $Q_{\rm w}$(erg g$^{-1}$ s$^{-1}$), for the three 
$\tau$'s (solid) with adiabatic cooling in eq. (\ref{eq:inteng}) (dashed). 
{\it Lower right} : Comparison of $\Gamma$.  
Solid lines are wave acceleration, $\Gamma_w$ ,dashed line is 
magneto-rotational acceleration, $\Gamma_b$, and dot-dashed line is 
acceleration by gas pressure, $\Gamma_g$. }
\label{fig:slmdrot}
\end{figure}

Figure \ref{fig:slmdrot} presents result of moderate magneto-rotational wind 
for $S_0 = 0.15$ in a solar-mass star ($M=1M_{\odot}$ and $R=1R_{\odot}$). 
We adopt $B_{r,0} = 8{\rm G}$ and $T=2\times 10^6$K as typical values. 
A larger mass loss rate, $\dot{M}_{\rm E} = 2\times 10^{-13} 
M_{\odot}$yr$^{-1}$, 
is employed than the present solar value, $\dot{M} \simeq 
2\times 10^{-14} M_{\odot}$yr$^{-1}$, because  
young solar-mass stars are found to have larger mass loss rate \citep{wod02, 
wod05} possibly as a result of their enhanced coronal activities or 
stellar rotation. 

On the left of fig \ref{fig:slmdrot},
we show the wind structures as a function of $r$. 
The top panel shows the velocity distributions, $v_r$ and $v_{\phi}$ 
of our chosen empirical model, and compares
them with both $v_f$ and the rigid rotation speed $r\Omega$.  
The Michel velocity is $v_{\rm M} = 1018$ km/s, and is selected as $v_{\infty}$
following eq. (\ref{eq:tmvl}). This value is larger than the escape speed, 
$v_{\rm esc,0}=617$ km/s. For the radial velocity
distribution, we adopt the velocity law index $\eta=3$ (in eq:\ref{eq:vrstr}), 
although we find that $\eta$  only weakly affects the results.  
It is the difference between $v_{\phi}$ and the solid body rotation,
$r\Omega$, leads to the spiral pattern.  
The middle panel shows that the angle $\theta$ increases monotonically up 
to $\gtrsim 70^{\rm o}$. 
The bottom panel presents the plasma $\beta$ value 
(eq. (\ref{eq:plsmbt})), and it shows that moderately low-$\beta$ 
circumstances are in fact realized for these calculations. 

Once we have chosen the empirical wind structure for a star, 
we can solve for the propagation of the fast waves
by using eq. (\ref{eq:vrdv}). On the right side of fig. \ref{fig:slmdrot}, 
results for wave periods $\tau=1,
10, 100$s. The top panel exhibits
variation of $\delta w$ (solid) in comparison with the no-dissipation case
(dashed). Although $\delta w$ decreases initially owing to the rapid increase of
$v_{\rm f}$ to satisfy the wave energy equation (\ref{eq:wveng}), farther out
it is amplified because of the  outward propagation through a decreasing density
region. 
Waves with shorter periods, $\tau$, suffer greater
damping per unit length, so that they dissipate nearer the star.

The middle-right panel of fig.\ref{fig:slmdrot} shows a role of the wave 
dissipation on the wind energetics.  
An equation describing variation of internal energy per unit mass, $e$, is 
written as 
\begin{equation}
v_r \frac{de}{dr} + \frac{a^2}{r^2}\frac{d}{dr}(r^2 v_r) - Q_{\rm w} = 0 ,
\label{eq:inteng}
\end{equation}
where the second term denotes the adiabatic loss rate and the third term is
the heating caused by the wave dissipation. In our current study we evaluate 
the second and third terms using the chosen empirical wind model.
We can then determine the magnitude of the heating term for different $\tau$ 
(solid) and  compare these with the  adiabatic cooling rate (dashed). 
The panel illustrates that the region heated by the waves depends on 
the value of $\tau$. Waves with the smaller values of $\tau$ heat
the inner regions.  Note that the heating exceeds the adiabatic cooling in the 
respective regions, and this indicates that the assumed 
temperature ($=2\times 10^6$K) could indeed be
maintained by the wave dissipation. In the case of actual stars,
the heating could be distributed more uniformly than shown in 
the figure, if we were to take into account a spectrum of  waves
and include thermal conduction as well, as will be discussed in (\S\ref{sec:lmd}). 

The lower-left panel of fig.\ref{fig:slmdrot} shows the wind dynamics. 
The equation of momentum density is 
\begin{equation}
\frac{d}{dr}(\frac{v_r^2 + v_\phi^2}{2}) + \frac{1}{\rho}\frac{dp_g}{dr} 
 + \frac{1}{\rho}\frac{dp_{\rm fw}}{dr} + \frac{G M_{\star}}{r^2} 
- \frac{d}{dr} (\frac{r\Omega B_r B_\phi}{4\pi \rho v_r}) = 0, 
\end{equation}
\citep{lc99}, and the second and third terms are accelerations by the
gas and the wave pressure, respectively. A fifth term accounts for
magneto-rotational acceleration and it consists of centrifugal force and
acceleration owing to the slinging effect of the rotating magnetic field. We
compare these three terms in ratios relative to the gravitational
acceleration, i.e. 

$$\Gamma_g = \left| \frac{1}{\rho}\frac{dp_g}{dr} / \frac{G M_\star}{r^2}
\right|, $$

$$\Gamma_w = \left| \frac{1}{\rho}\frac{dp_{\rm fw}}{dr} / 
\frac{G M_\star}{r^2} \right| , \;\;$$

$$ \Gamma_b = \left| \frac{d}{dr} (\frac{r\Omega B_r B_\phi}{4\pi \rho v_r}) 
/ \frac{G M_\star}{r^2} \right| , $$ 
The magnitude of these terms has been evaluated using the empirical wind 
model and the results are shown in the panel. 
Note that the gas pressure and the magneto-rotational force are almost 
comparable in the outer region, $r\gtrsim 10R_\star$. This shows that it is 
reasonable to call the wind ``a moderate magneto-rotator''. The figure 
also illustrates that the wave pressure becomes as effective an
acceleration term as these other two components for a magneto-rotational wind.
This means that the fast wave damping is equally important in the wind dynamics; 
thus the wave pressure could further accelerate the wind.  

\begin{figure}
\figurenum{5} 
\epsscale{1.} 
\plotone{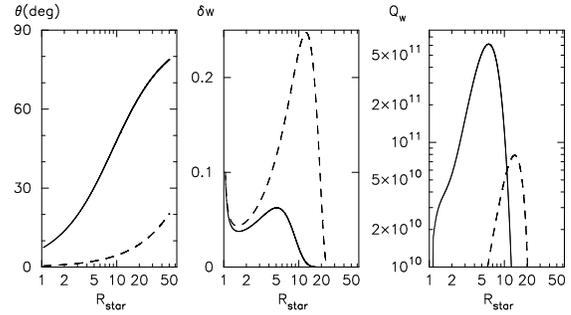} 
\caption{Comparison of the propagation angle $\theta$, amplitude $\delta w$,
and heating rate $Q_{\rm w}$(erg g$^{-1}$ s$^{-1}$) for a fast and slow 
rotating solar-mass star.
The solid lines show results for the fast rotator, $S_0 = 0.5$,
and the dashed lines refer to the slow rotator, $S_0 =0.01$. Note that
the angle and heating rates for the fast rotator are larger, but
the wave amplitude (middle panel) for the fast rotator is smaller. This
is because it is the radial decrease in the amplitude that leads to the
larger heating shown in the right panel.
}
\label{fig:slmdfs}
\end{figure}

Next, we study effects of rotation rate. In fig. \ref{fig:slmdfs}, we
compare cases of slow rotator ($S_0=0.01$) and fast rotator ($S_0 = 0.5$).
The angle $\theta$ is greatly different in both cases as illustrated in a
left panel. Accordingly, the waves dissipate more close to the star for the fast
rotator (a middle panel), and the heating (and acceleration not shown) for
the fast rotator occurs in the inner region (a right panel). 
As a result, the
wave energy effectively dissipates in the case of the fast rotator to heat and
accelerate the plasma, while  for the slow rotator,
the contribution is not as large. 
This indicates that regions heated by the 
wave dissipation depend on the stellar rotation even when the same waves
amplitudes are incident at the base.

As a star evolves, the spin rate $S_0$ generally decreases. Using this tendency,
we can speculate on how stellar evolution affects the wave dissipation.
Observationally, it is  known that young solar-types stars 
have both a more rapid rotation and show a stronger X-ray luminosity
than is the case for the sun \citep{piz03}. This indicates
that some processes are operating to heat the coronae in the 
younger, more rapidly rotating stars, in addition
to the ``basal'' heating processes occurring in the present solar corona.
We suggest that the process of collisionless damping of fast MHD 
waves can play an  important role in the enhanced heating of the coronae and
winds and the X-ray source regions of  young solar-type stars because 
it is much more effective in the rapidly rotating young stars.

\subsection{Massive Stars}
We have shown that our process possibly works in low mass main sequence stars 
with moderately fast rotation. In this subsection, we would like to consider 
cases of massive stars. 
Traditionally, massive main sequence stars have not been thought to have
activities concerning magnetic fields, i.e. coronae and/or flares, because
their interiors have an outer radiative zone which impedes the rise of
magnetic fields and there was no known dynamo mechanism that would
lead to surface fields.  Nevertheless, as we have stated in \S
\ref{sc:intro}, there has been increasing observational evidence 
for surface fields on hot stars. Coronal activities have been considered
to explain the X-ray properties of mid to late B-type near-main-sequence stars
\citep{ccm97}. More recently, \citet{mc03} have shown that the massive stars 
can have a surface field as a result of a buoyant rise through the 
radiative zone of fields generated at core-envelope interface dynamo.
Although the coronal like geometry might be rather different from 
the less massive stars, we consider it interesting and instructive
to apply the same basic model to massive stars.

\begin{figure}
\figurenum{6} 
\epsscale{.9} 
\plotone{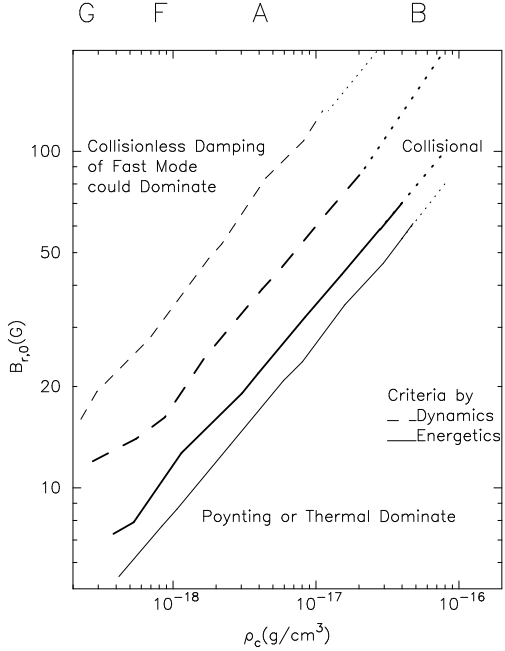} 
\caption{Critical magnetic field for collisionless damping of the fast waves 
as a function of characteristic density, $\rho_{\rm c}$ 
for spin parameter $S_0=0.15$ (thick) \& 0.5 (thin). 
We here consider the waves with 
$\tau=1$(s) and $\delta w_0=0.1$. Solid lines indicate criteria based on 
the energetics and dashed lines are those of the dynamics. 
Upper characters on the top denote corresponding main sequence stars 
(see text). 
In upper left region on each line the wave dissipation gives larger 
contributions to the heating or acceleration than the other processes. 
In upper right region which is indicated as 'Collisional', the plasma 
is still collisional even at $r = 2R_{\star}$ due to high density so that 
our process does not work effectively.}
\label{fig:otst}
\end{figure}

It is known from observations of spectral lines that early type
stars are commonly fast rotators. This is because such a star tends
to be rotating quickly after its initial collapse and formation, the
lifetimes are rather short, and thus the massive stars are able to 
persist in having rapid rotation. 
The observed mass-loss rate, $\dot{M}_{\rm obs}$, of these types of stars is 
known for given $M_\star$ and $R_\star$ \citep{jnh88}. 
Generally, $\dot{M}_E \gtrsim \dot{M}_{\rm obs}$ in the magneto-rotational 
winds since equatorial mass flux is larger than polar mass flux.  
Magnetic activities and/or stellar rotation might further enhance intrinsic 
$\dot{M}$ as discussed in the previous subsection. 
To take into account these effects, we consider 
$\dot{M}_E \simeq (1-10)\dot{M}_{\rm obs}$.
Then, we can construct the wind structure 
for a certain star after fixing $B_{r,0}$ and $S_0$ (\S \ref{sc:apwdst}). 
Here, we would like to introduce a characteristic mass-loss 
density \citep{cbm02} defined as    
\begin{equation}
\label{eq:crdns}
\rho_{\rm c} \equiv \frac{{\dot M}_{\rm E}}{4 \pi R_{\star}^2 v_{\infty}}. 
\end{equation}
We have found that $\rho_{\rm c}$ is a good indicator for our process; 
the damping of the fast waves work similarly in  
winds with the same $\rho_{\rm c}$ but different $M_\star$, $R_\star$, 
and $\dot{M}_E$. 
This is because the damping is controlled by density in the winds by way
of the  plasma $\beta$ value which appears in eq.(\ref{eq:damp}) for 
the damping rate.  

We now examine the roles of the fast waves in the wind energetics and 
dynamics for three stellar parameters, $\rho_{\rm c}$, $B_{r,0}$ and $S_0$. 
Since our process does not work dominantly in the slow rotators because
the value of $\theta$ is 
too small in the winds, we study cases with $S_0=0.15$ (moderate 
rotator) and $0.5$ (fast rotator). We investigate the wave propagation 
in main sequence stars with $M_\star = 1 - 8M_{\odot}$. $R_\star$ for 
a given $M_\star$ is determined by a relation for the main sequence stars 
\citep{kw90}. Temperature is assumed to be $T=2\times 10^6$K, independent of 
$M_\star$. 
Several cases of $\dot{M}_{\rm E}$ for a fixed $(M_\star,R_\star)$ 
are calculated within the range of $\dot{M}_{\rm E}=(1-10)M_{\rm obs}$. 
We consider waves with $\tau=1$s and $\delta w_0=0.1$ and investigate 
how the wave dissipation affects the energetics and dynamics 
for different $B_{r,0}$. 

As for the energetics, we examine whether the heating integrated with 
respect to a certain region exceeds the adiabatic cooling integrated in the 
same region, namely, we determine the condition of $B_{r,0}$ which 
satisfies    
\begin{equation}
\label{eq:engtc}
\int_{R_\star}^{r_{\rm out}} dr \rho r^2 Q_{\rm w} \ge 
\int_{R_\star}^{r_{\rm out}} dr \rho a^2 \frac{d}{dr}(r^2 v_r).  
\end{equation}
We adopt $r_{\rm out}=10R_\star$ to focus on the heating in the inner 
wind region. 
We also carry out the similar procedure for the dynamics. 
We compare the acceleration due to the wave pressure integrated in 
the same region with the acceleration by the other two components: 
$$
-\int_{R_\star}^{r_{\rm out}} dr v_r r^2 \frac{dP_{\rm w}}{dr} 
\ge -\int_{R_\star}^{r_{\rm out}} dr v_r r^2 \frac{dp}{dr}  
$$
\begin{equation}
\label{eq:dynmc}
+ \int_{R_\star}^{r_{\rm out}} dr \rho v_r r^2  
\frac{d}{dr} (\frac{r\Omega B_r B_\phi}{4\pi \rho v_r})
\end{equation}
  
Figure \ref{fig:otst} shows the critical values for $B_{r,0}$ with respect to 
the energetics 
(solid; eq. (\ref{eq:engtc})) and dynamics (dashed; eq. (\ref{eq:dynmc})) 
as a function of $\rho_{\rm c}$ in the moderate (thick) and fast (thin) 
rotators. 
The collisionless damping of the fast modes can dominate the other processes 
if the field strength exceeds the respective lines. 
If $\dot{M}_{\rm obs}$ is used when deriving $\rho_{\rm c}$ instead 
of $\dot{M}_{\rm E}$ in eq. (\ref{eq:crdns}), $\rho_{\rm c}$ has one-to-one 
correspondence to $M_\star$ because dependence of $\dot{M}_{\rm obs}$ 
on $M_\star$ is steep \citep{jnh88} while those of $R_{\star}$ and 
$v_{\infty}$ are weaker \citep{kw90}. 
At the top of the figure, we give the  spectral type of the stars 
used to determine $\rho_{\rm c}$. 
Please note that this is based on $\dot{M}_{\rm obs}$. 
If a star has larger $\dot{M}_{\rm E}$ as discussed previously, 
its $\rho_{\rm c}$ becomes larger than the shown spectral type.

Figure \ref{fig:otst} shows that the required value for $B_{r,0}$ 
is $\lesssim 100$G even in mid to late B stars. 
This is a reasonable range for these types of stars 
as has been shown by Maheswaran \& Cassinelli (1992). Cassinelli \etal
(2002) have shown that such fields are sufficient to play a role in the 
channelling and torquing of the moderately weak winds of B main-sequence 
at about spectral type B3V and later. 
The critical $B_{r,0}$ is a monotonically increasing function of 
$\rho_{\rm c}$ and a relation, $B_{r,0}^2/\rho_{\rm c} \sim$const, holds.
Since we assume the same gas temperature, the relation can be understood 
that the similar plasma $\beta$ is required even for different stellar mass. 
As $\rho_{\rm c}$ increases, the collisionless condition breaks 
down in the inner regions. On the dotted lines the plasma is still collisional 
at $r=2 R_\star$ for the waves with $\tau=1$s. From the magnetic fields, of
order 100 Gauss or so, that are required, we can conclude that 
our process can operate effectively in stars with spectral type of about
B3V  and later with hot outer atmospheres/winds.
Interestingly, this spectral type is consistent with the result, 
obtained by \citet{ccm97},  that the X-rays from mid- to late-  B type 
stars seem to be arising increasingly from the presence of extensive hot 
winds and confined coronal zones. 

In both cases of $S_0=0.15$ and 0.5, the condition that the damping
dominate the dynamics demands a larger $B_{r,0}$ than is the case
for the energetics. This is because the dynamics condition 
compares the wave acceleration with both thermal and Poynting processes 
while the energetics condition compares the damping with the thermal 
processes (i.e. adiabatic cooling) only.  
In a region between the two conditions, the wind could be heated by the 
fast waves, but still accelerated dominantly by the magneto-rotational force 
rather than by the waves. 
For the moderate rotators, the wave dissipation is slower and a certain 
fraction of the wave energy remains at the radius $r_{\rm out}$. 
Therefore, the magnetic field  $B_{r,0}$ that is required 
for the energetics condition is larger 
than that for the fast rotators, and this is to increase the initial 
wave energy. 
On the other hand, the dynamics condition shows that the fast rotators 
demand the larger $B_{r,0}$, and this is to let the wave acceleration 
exceed the magneto-rotational acceleration, which is also larger for the
fast rotators.

\section{Discussion}
So far we have studied outward propagation of the fast waves based on 
several assumptions : (i) wave amplitude at the surface 
($\delta w_0=0.1$) (ii) monochromatic wave spectrum (iii) no refraction 
(iv) a fixed, pre-specified  magneto-rotational wind. 
First, we discuss these assumptions. 
Then, we compare the collisionless damping with the steepening damping
and with other types of waves as well. 
Finally we mention another application of our process, to pulsar wind nebulae.

\subsection{Limitations of Model}
\label{sec:lmd}
\hspace{-1cm}(i) {\it ~Wave Generation} :
In \S\ref{sc:rslt} we have assumed that $\delta w_0 = 0.1$ at the inner 
boundary without specifying processes of the wave generation. 
The inner boundary coincides with not the photosphere but the base of 
the corona, since we assume $T=2\times 10^6$ throughout the atmosphere. 
The density generally decreases by several orders of magnitude from the 
photosphere to the corona, and the region between them is called chromosphere 
and trangition region. 

We can consider two types of the wave generators. 
One type of wave generator is associated with the granulations originating 
from the surface convection zones which exist in the low- to intermediate 
stars. The other type of generator corresponds to  oscillations of 
closed magnetic loops emerging from the surface. The up-going waves generated 
by the 
former process suffer damping in the chromosphere before reaching the corona 
(see Stein Schwartz 1972 for sound waves).  
The closed magnetic loop process directly excites the waves at higher 
altitude in the corona \citep{str99}, and thus is more suitable for the 
present paper. 
Stars with faster rotation are expected to have stronger activities involving 
these magnetic loops \citep{jar02}.    

Accordingly, $\delta w_0$ would be larger in these stars than in stars 
with slower rotation; probably $\delta w_0=0.1$ is realizable in the 
fast and moderate rotators by the direct excitation in the corona, while 
it is less likely in the slow rotators due to insufficient 
activity by the loops. In addition to the difference of the dissipation 
conditions (\S\ref{sc:rslt}), the difference of the wave generation processes 
would be important when discussing the stronger X-ray activities in faster 
rotating stars. 

\hspace{-1cm}(ii) {\it ~Wave Spectrum} : 
Although we have only considered the monochromatic waves in the model 
calculations, they are likely to be injected from actual stars 
with some spectrum of frequencies, and the wave frequency controls the wave
damping as we have shown in the previous section. 
Namely waves with higher frequencies dissipate in the inner region. 
If some spectrum (\eg\ a power-law in $\omega$) were given for the input waves, 
the stellar atmosphere would be
heated more uniformly; with the inner regions is heated by waves with higher 
frequencies, and outer regions is heated by those with lower frequencies.

\hspace{-1cm} (iii){\it ~Wave Refraction} :
We have assumed that the fast mode propagates radially without being
refracted. In fact, the refraction in not negligible unless the plasma
$\beta$ is sufficiently small. This is because the anisotropy of $v_{\rm f}$
appears in cases in which the sound speed is not sufficiently small relative
to the \Alfven speed (eq. (\ref{eq:fstwsp})). Generally, a wave is refracted
toward a region with a smaller phase speed. For a fast mode wave, it is
refracted in a direction away from the field line (toward larger $\theta$).
This effect causes the fast wave to dissipate more quickly since the damping
increases with the angle $\theta$. As a result, we should expect the heated region to
be shifted slightly inward as compared with the heated region
from the radial propagation that we assumed.

\hspace{-1cm} (iv){\it ~Fixed Background} :  
In this paper, we treat the wave propagation within a given empirical wind 
structure.  We think that this approach is acceptable as a first attempt to 
assess the importance of heating and acceleration produced by wave dissipation. 
Needless to say, a self-consistent treatment of the wind structure
would be required in a more complete modeling of a star. This is because
wave dissipation can directly affect both $v_r$ 
(via acceleration) and $T$ (via heating). These two variables affect
other quantities associated with the outflow as well.
However, we are of the opinion that empirical $v_r$ that we have
assumed gives reasonable wind structure, simply because the 
velocity distribution  is determined mainly by the basic 
stellar  parameters $M_{\star}, R_{\star}, S_0$. Figure \ref{fig:otst}
also illustrated that a stronger field was needed to modify the 
dynamics than is the case for the energetics.

Given the velocity structure, it is the energetics and the 
resultant temperature that must be considered  more carefully. 
Unlike the wind velocity, the temperature structure
is not determined by the basic stellar parameters but 
rather by the boundary conditions for the wave generation and the 
subsequent wave heating and radiative cooling and thermal  conduction
cooling . 
An important point for our collisionless process is that the damping rate 
is controlled by temperature through $\beta$ value (eq. (\ref{eq:damp})). 
Once the plasma is heated, i.e. $\beta$ increased, by the waves, the damping 
is enhanced. It leads to further heating until most of the wave energy 
dissipates. This seemingly catastrophic behavior is partially inhibited 
in reality by the cooling effects of thermal conduction which 
become more effective if the temperature increases.  
In more intensive studies, it would be important to examine
the more complicated energetics with the other effects are accounted for.

\subsection{Collisionless Damping vs. Steepening}

We would like to compare the two types of the damping discussed earlier,
in light of the magneto-rotational wind results. 
Figure \ref{fig:damp} shows that the collisionless 
process and steepening work in different regimes, namely the former is 
important for larger $\theta$ except $\theta = 90^{\rm o}$ and smaller 
$\delta w \lesssim 0.1$ while the latter dominantly works for smaller 
$\theta$ and larger $\delta w > 0.1$. 

In a stellar atmosphere the amplitude, $\delta v$, of the upgoing wave is 
inevitably amplified because the density decreasing. 
Even the normalized amplitude, $\delta w$, increases after the slight decrease 
in the inner region as shown in the top right panel of 
Figure \ref{fig:slmdrot}. 
In the cases of moderate and fast rotators, $\theta$ is already large when 
$\delta w$ becomes large enough for the dissipation. 
Therefore, the collisionless damping sets in first. 
$\delta w$ is kept small by the 
collisionless damping in the entire region, which is not suitable 
for the steepening.  In the case of slow rotators, the collisionless 
damping is not so effective even though $\delta w$ increases to the same 
level, because $\theta$ is still small.   
As a result, the dissipation of waves through shock steepening is expected 
to be more important in slow rotating stars.

\subsection{Other Modes}
\label{sc:otmd}
In this paper we have focused on the fast mode waves. 
In reality, the three MHD modes couple and have a more-or-less turbulent nature. 
Since the waves and turbulence are mainly excited from the stellar surface, 
one has to consider the imbalanced cascade of turbulence
in such a situation \citep{clv02}.    
Also, because of the density stratification, the amplitude of the turbulence 
would not decrease with radius even in the outer region, far from the 
surface. 
  
The slow wave has not been considered as a dominant heating source in the 
solar corona, because the slow waves that are generated from the photosphere 
dissipate too rapidly \citep{ss72}. 
However, it has been shown that
strong driving of the slow mode could 
occur at higher altitude as a result of the dynamical motion of the closed loops 
\citep{str99} in young solar type-stars that exhibit strong stellar 
activities \citep{ayr00}. In such a situation, supersonic slow modes can
be important \citep{blc05}. 
Slow mode waves suffer TTD, similar to that affecting fast mode waves, and 
thus contribute to the heating. This will be studied in a forthcoming
paper. The steepening would be important as well \citep{suz02}. 

The \Alfven wave is also a  candidate for the heating 
in the low-$\beta$ solar and stellar plasmas. 
The \Alfven waves dissipate in density-stratified atmosphere by processes
such as phase mixing \citep{hp83}, nonlinear mode conversion \citep{cl02,cl03,
si05}, turbulent cascade triggered by reflection \citep{oug01}. 
High frequency \Alfven waves suffer ion-cyclotron damping \citep{cfk99} as well. 
The dissipative nature of the \Alfven waves, which propagate along the field 
lines in the magneto-rotational winds is less modified for cases in which
the wave length is shorter than the curvature scale of the winding field 
lines. So, \Alfven waves can operate in both slow and fast rotators 
and  also contribute to the basal  heating in stellar coronae. 
The processes directly linking 
to the stellar rotation as shown in this paper would give additional 
contributions to lead to stronger coronal activities in rapidly rotating 
stars.


\subsection{Application to Pulsar Winds}

So far we have focused on the collisionless damping of the fast MHD waves in 
normal stars. This mechanism can work in any situation in which 
the fast waves travel in a low $\beta$ plasma with curved magnetic fields. 
A pulsar wind is one of the candidates. A pulsar magneto-sphere 
is a low $\beta$ and collisionless plasma, and the MHD approximation is 
acceptable because the particles can interact with the waves
through the magnetic fields. 

\citet{lyu03} considered dissipation of the fast waves propagating radially 
in pulsar winds by multiple shocks as a result of the steepening in order 
to investigate dissipation of the electromagnetic energy. 
Because the toroidal magnetic field component dominates owing to  
the rotation of the pulsars, 
he studied the fast waves traveling perpendicular to the field lines, i.e. 
$\theta = 90^{\circ}$. 
For this specific angle, the collisionless damping is not efficient,
as shown in \S \ref{sc:dampfs}, 
so that the steepening would in fact be the more 
effective process in the wave dissipation. However, if $\theta$ is not
strictly $=90^{\circ}$, say because of perturbed magnetic fields or other
effects,  collisionless damping would again become important. 
Although beyond the scope of the present paper,
relativistic effects should be taken into account
in a proper study of collisionless damping in pulsar winds.

\section{Summary}

For the case of magneto-rotationally driven winds, we have considered the 
heating and acceleration of a stellar plasma caused by
the collisionless damping of fast MHD waves. 
Fast mode waves propagate almost radially even in spiral magnetic 
fields thanks to the isotropic character of the propagation, provided that 
the $\beta$ is low. 
The plasmas are typically collisionless, so the waves undergo  
collisionless damping. We find that collisionless damping of fast MHD waves
is effective when 
(i) the angle between the wave propagation and the field line is large and
(ii) the plasma $\beta$ is around $0.01 - 0.1$. 
In the magneto-rotational winds, the angle increases 
as the wave travels away from the surface. This angle change
eventually leads to a dissipation 
of the waves, and causes heating and acceleration of the surrounding medium.

We have found that our collisionless damping is important in 
moderate and fast magneto-rotators of solar-mass stars, 
while it is 
not so effective in slow rotators such as the (present) sun. 
It cooperates with the ``basal'' processes which are responsible for the 
heating of the corona in the slow rotators. We propose that the
mechanism of collisionless damping of fast waves can explain the observed 
strong 
coronal activities in young (age $\lesssim 100$Myr) solar-type stars 
because of their rapid rotation.  
The damping also effectively applies in the energetics, and perhaps also to the
dynamics, of the atmospheres of moderately massive main sequence stars of which 
are of spectral types about B3V or later and which have observationally plausible 
magnetic fields ($\lesssim 100$G). 

When compared with the shock dissipation by the steepening, the collisionless 
damping is expected to be more important in moderate and fast rotators with 
an input wave amplitude, $\delta w_0 \lesssim 0.1$. 
In addition, the process is also argued to be important in general 
astrophysical low $\beta$ plasma which has curved magnetic fields, such as 
pulsar magneto-spheres.

T.K.S. is supported by the JSPS Research Fellowship for Young Scientists, 
grant 4607, and acknowledge a Grant-in-Aid for the 21st Century COE 
``Center for Diversity and Universality in Physics'' from the Ministry of 
Education, Culture, Sports, Science, and Technology (MEXT) of Japan.
A.L. and H.Y. acknowledge NSF grant ATM-0312282, and A.L. and J.P.C. 
also acknowledge the 
support of the Center for Magnetic Self-Organization in Laboratory and 
Astrophysical Plasmas.

\appendix
\section{Damping by Steepening}
\begin{figure}
\figurenum{7} 
\epsscale{.4} 
\plotone{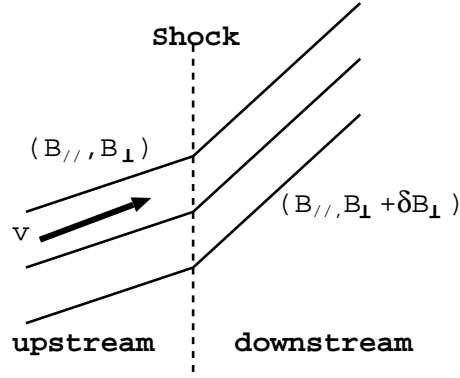} 
\caption{Geometry of fast shock. }
\label{fig:fsshgm}
\end{figure}

In this appendix, we summarize the damping of fast shock trains which form
by the steepening of fast MHD waves. First, we derive entropy generation at
the fast shock whose geometry is shown in fig. \ref{fig:fsshgm}. Physical
quantities in the upstream region are indicated by subscript ``1'' and those
in the downstream region by ``2''. Components parallel with shock normal are
denoted by subscript ``$\parallel$'' and those perpendicular to it by
``$\perp$''. Because of the conservation of magnetic flux, $\mbf{\nabla
\cdot B} = 0$, the parallel component of magnetic field, $B_{\parallel}$, is
constant. The perpendicular field strength is increased from $B_{\perp}$ to
$B_{\perp} + \delta B_{\perp}$ at the fast shock. The equations for the
conservation of mass flux, momentum parallel and perpendicular to shock
normal, electric field, and energy are expressed in the frame of
an observer co-moving with the shock as (e.g. Priest \& Forbes 2000)
\begin{equation}
\label{eq:fshms}
\rho_1 v_{\parallel,1} = \rho_2 v_{\parallel,2}, 
\end{equation}
\begin{equation}
\label{eq:fshmpr}
p_1 + \rho_1 v_{\parallel,1}^2 = p_2 + \rho_2 v_{\parallel,2}^2 
+ \frac{2 B_{\perp}\delta B_{\perp} + \delta B_{\perp}^2}{8\pi},
\end{equation}
\begin{equation}
\rho_1 v_{\parallel,1} v_{\perp,1} = \rho_2 v_{\parallel,2} v_{\perp,2} 
- \frac{B_{\parallel} \delta B_{\perp}}{4\pi},
\end{equation}
\begin{equation}
\label{eq:fshef}
v_{\parallel,1}B_{\perp} - v_{\perp,1}B_{\parallel} = v_{\parallel,2}
(B_{\perp} +\delta B_{\perp}) - v_{\perp,2}B_{\parallel} (=0)
\end{equation}
and
$$
\hspace{-5cm}
p_1 v_{\parallel,1} + \rho_1 e_1 v_{\parallel,1} + \frac{B_{\perp}^2}{4\pi}
v_{\parallel,1} - \frac{B_{\parallel}B_{\perp}}{4\pi} v_{\perp,1} 
+ \frac{1}{2}\rho_1 (v_{\parallel,1}^2 + v_{\perp,1}^2) v_{\parallel,1} 
$$
\begin{equation}
\label{eq:fsheg}
= p_2 v_{\parallel,2} + \rho_2 e_2 v_{\parallel,2} + \frac{(B_{\perp} 
+\delta B_{\perp})^2}{4\pi}v_{\parallel,2} - \frac{B_{\parallel} (B_{\perp} 
+\delta B_{\perp})}{4\pi} v_{\perp,2} 
+ \frac{1}{2}\rho_2 (v_{\parallel,2}^2 + v_{\perp,2}^2) v_{\parallel,2}.  
\end{equation}
Here we define all the quantities in a frame co-moving with the shock front. 
Then, both left- and right-hand sides of eq.(\ref{eq:fshef}) become 0. 
By eliminating $v_{\parallel,1}$, $v_{\parallel,2}$, $v_{\perp,1}$,  
$v_{\perp,1}$, and $p_2$ from 
eqs.(\ref{eq:fshms})--(\ref{eq:fsheg}), we have  
a relation 
between $\delta B_{\perp}$ and compression ratio, 
\begin{equation}
\sigma\equiv \frac{\rho_2}{\rho_1}, 
\end{equation}
as 
\begin{equation}
\label{eq:fshdbs}
-\frac{\gamma p_1}{\rho_1}\frac{\sigma -1}{\sigma} 
- \frac{\delta B_{\perp}^2}{8\pi \sigma \rho_1}
+ \{\gamma + 1 - \sigma (\gamma -1)\} \frac{\delta B_{\perp}}
{8\pi\sigma\rho_1}\frac{\{(\sigma -1) B^2 - B_{\perp}\delta B_{\perp}\}}
{\{\delta B_{\perp} - (\sigma -1)B_{\perp}\}}=0, 
\end{equation}
where $B^2 = B_{\parallel}^2 + B_{\perp}^2$. 
The pressure jump is also derived from eq. (\ref{eq:fshmpr}) :
\begin{equation}
\label{eq:pjump}
\Delta p \equiv p_2 -p_1 
= \frac{2(\sigma-1)B^2 \delta B_{\perp} + (\sigma -3) B_{\perp} \delta 
B_{\perp}^2 - \delta B_{\perp}^3}{8\pi \{\delta B_{\perp} - 
(\sigma -1)B_{\perp}\}}. 
\end{equation}
Entropy generation is written as 
\citep{mm84}
\begin{equation}
\label{eq:entrp}
\frac{\Delta s}{R}  = \frac{1}{\gamma -1}\ln(\frac{p_2 \rho_1^\gamma}
{p_1 \rho_2^\gamma})
\simeq \frac{1}{\gamma -1} \{ \ln(1 + \frac{\Delta p}{p}) 
- \gamma \ln(\sigma)\}, 
\end{equation}
where $R$ is a gas constant, and we assume that the shock is weak. 
Then, $\Delta s$ is derived from a given amplitude, $\delta B_{\perp}$, by 
using eqs.(\ref{eq:fshdbs}) and (\ref{eq:pjump}).

The steepening of the wave fronts forms the fast shock train. The wave 
dissipates by these multiple shocks.  
The heating rate, $Q_{\rm s}({\rm erg\; cm^{-3}s^{-1}})$, of the fast shock train 
with angular frequency, $\omega$, is given by 
\begin{equation}
Q_{\rm s} = \frac{\Delta s}{R} \rho_1 R T \frac{\omega}{2\pi} = 
\frac{\Delta s}{R} p_1 \frac{\omega}{2\pi}. 
\end{equation}
The damping rate is derived from $Q_{\rm s}$ and energy density of the wave, 
$\frac{\delta B_{\perp}^2}{8\pi}$, as 
\begin{equation}
\label{eq:dmpap}
\gamma_{\rm s} = \frac{Q_{\rm s}}{\delta B_{\perp}^2/8\pi}
\end{equation}
Note that if $\delta B_{\perp}/B$ is fixed for the weak shocks, 
$\gamma_{\rm s}$ is independent of plasma $\beta$ because 
\begin{equation}
\label{eq:dmpscl}
\gamma_{\rm s} \propto \frac{\Delta s}{R} p_1 \frac{1}{\delta B_{\perp}^2} 
\simeq \frac{\Delta p}{p_1} p_1 \frac{1}{\delta B_{\perp}^2} 
\propto B^2\frac{1}{\delta B_{\perp}^2}
\end{equation}
from eqs.(\ref{eq:pjump}) --- (\ref{eq:dmpap}). 


\begin{thebibliography}

\bibitem[Ayres (2000)]{ayr00}
Ayres, T. R. 2000, \solphys, 193, 273

\bibitem[Barnes(1966)]{bar66}
Barnes, A. 1966, Phys. Fluid, 9, 1483

\bibitem[Barnes(1968)]{bar68}
Barnes, A. 1968, \apj, 154, 751

\bibitem[Barnes(1969)]{bar69}
Barnes, A. 1969, \apj, 155, 311

\bibitem[Barnes (2003)]{bar03}
Barnes, S. A. 2003, \apj, 586, 464

\bibitem[Belcher (1971)]{bel71}
Belcher, J. W. 1971, \apj, 168, 509

\bibitem[Belcher \& MacGregor(1976)]{bm76}
Belcher, J. \& MacGregor, K. B. 1976, \apj, 210, 498

\bibitem[Beresnyak, Lazarian, \& Cho(2005)]{blc05}
Beresnyak, A., Lazarian, A., Cho, J. 2005, \apjl, 624, L93

\bibitem[Bergh\"{o}fer et al.(1997)]{ber97}
Bergh\"{o}fer, T. W., Schmitt, J. H. M. M., Danner, R., \& Cassinelli, J. P. 
1997, \aap, 322, 167


\bibitem[Cassinelli et al.(2002)]{cbm02} 
Cassinelli, J. P., Brown, J. C., Maheswaran, M., Miller, N. A., \& 
Telfer, D. C. 2002, \apj, 578, 951

\bibitem[Cassinelli \& Olson (1979)]{co79} 
Cassinelli, J. P. \& Olson, G. L. \apj, 1979, 229, 304

\bibitem[Cassinelli \& Swank (1983)]{cs83} 
Cassinelli, J. P. \& Swank, J. H. \apj, 1983, 271, 681

\bibitem[Charbonneau \& MacGregor (1980)]{cm80}
Charbonneau, P., \& MacGregor, K. B. 2001, \apj, 559, 1094

\bibitem[Cho \& Lazarian(2002)]{cl02}
Cho, J. \& Lazarian, A. 2002, \prl, 88, 245001

\bibitem[Cho, Lazarian, \& Vishniac(2002)]{clv02}
Cho, J., Lazarian, A., Vishniac, E., T. 2002, \apj, 564, 291

\bibitem[Cho \& Lazarian (2003)]{cl03}
Cho, J. \& Lazarian, A. \mnras, 2003, 345, 325

\bibitem[Cho, Lazarian, \& Vishniac(2003)]{clv03}
Cho, J., Lazarian, A., Vishniac, E., T. 2003, Lec. Not. in Phys., 614, 56

\bibitem[Cohen, Cassinelli, \& MacFarlane (1997)]{ccm97}
Cohen, D. H., Cassinelli, J. P. \& MacFarlane, J. J. 1997, \apj, 487, 867

\bibitem[Cranmer, Field, \& Kohl (1999)]{cfk99}
Cranmer, S. R., Field, G. B., \& Kohl. J. L. 1999, \apj, 518, 937 

\bibitem[Donati etal (2001)]{do01}
Donati, J.-F., Wade, G.A., Babel, J., Henrichs, H.F., de Jong, J.A., \&
Harries, T.J. 2001, MNRAS 326, 1265

\bibitem[Ginzburg (1961)]{gin61}
Ginzburg, V. L. 1961, {\it Propagation of Electromagnetic Waves in Plasma}, 
New York: Gordon \& Breach

\bibitem[Hartmann \& MacGregor (1980)]{hm80}
Hartmann, L. \& MacGregor, K. B. 1980, \apj, 242, 260

\bibitem[Henrichs (2001)]{hen01}
Henrichs, H.F. 2001, in {\it Magnetic Fields across the Hertzsprung-Russell
Diagram} ASP Conf Ser. Vol 248, 393

\bibitem[Heyvaerts \& Priest(1983)]{hp83}
Heyvaerts, J. \& Priest, E. R. 1983, \aap, 117, 220 

\bibitem[Hollweg (1982)]{hol82}
Hollweg, J. V. 1982, \apj, 254, 806

\bibitem[Jacques(1977)]{jaq77}
Jacques, S. A. 1977, \apj, 215, 942

\bibitem[de Jager, Nieuwenhuijzen, \& van der Hucht(1988)]{jnh88} 
de Jager, C., Nieuwenhuijzen, H., \& van der Hucht, K. A. 1988, \aaps, 72, 259

\bibitem[Jardine et al.(2002)]{jar02}
Jardine, M., Wood, K., Collier Cameron, A., Donati, J.-F., Mackay, D. H. 
2002, \mnras, 336, 1364

\bibitem[Kippenhahn \& Weigert (1990)]{kw90}
Kippenhahn, R. \& Weigert, A. 1990, ``Stellar Structure and Evolution'', 
{\it Springer-Verlag}

\bibitem[Lamers \& Cassinelli(1999)]{lc99}
Lamers, H. J. G. L. M. \& Cassinelli, J. P. 1999, 'Introduction to 
Stellar Wind', Cambridge

\bibitem[Lamers \& Morton(1976)]{lm76}
Lamers, H. J. G. L. M. \& Morton, D. C. \apjs, 1976, 32, 715

\bibitem[Lazarian \& Vishniac (1999)]{lv99}
Lazarian, A. \& Vishniac, E. T. 1999, \apj, 517, 700

\bibitem[Leamon et al.(1998)]{lmsw98}
Leamon, R. J., Matthaeus, W. H., Smith, C. W., \& Wong, H. K. 1998, \apj, 
507, L181

\bibitem[Lucy (1982)]{luc82}
Lucy, L. B. 1982, \apj, 255, 286

\bibitem[Lyubarsky (2003)]{lyu03}
Lyubarsky, Y. E. 2003, MNRAS, 339, 765

\bibitem[MacGregor \& Cassinelli (2003)]{mc03}
MacGregor, K. B. \& Cassinelli, J. P. 2003, \apj, 586, 480

\bibitem[Maheswaran \& Cassinelli (1992)]{mc92} 
Maheswaran, M. \& Cassinelli, J. P. 1992, \apj, 386, 695

\bibitem[Michel(1969)]{mc69}
Michel, F. C. 1969, \apj, 158, 727

\bibitem[Mihalas \& Mihalas (1984)]{mm84}
Mihalas, D. \& Mihalas, B. W. 1984, {\it Foundation of Radiation 
Hydrodynamics, Oxford University Press}, section 56  

\bibitem[Oughton et al.(2001)]{oug01} 
Oughton, S., Matthaeus, W. H., Dmitruk, P., Milano, L. J., Zank, G. P., 
\& Mullan, D. J. 2001, \apj, 551, 565


\bibitem[Parker (1963)]{pak63}
Parker, E. N. 1963, '{\it interplanetary Dynamical Processes}', John Wiley and 
Sons, New York

\bibitem[Parker (1966)]{pak66}
Parker, E. N. 1966, \apj, 145, 811


\bibitem[Pizzolato et al.(2003)]{piz03}
Pizzolato, N., Maggio, A., Micela, G., Sciortino, S., \& Ventura, P. 2003, 
\aap, 397, 147

\bibitem[Priest \& Forbes (2000)]{pf00}
Priest, E. \& Forbes, T. 2000, {\it Magnetic Reconnection}, Cambridge

\bibitem[Seward et al.(1979)]{sfg79}
Seward, F. D., Forman, W. R., Giacconi, T., Griffiths, R. E., 
Harnden, F. R., Jr., Jones, C., \& Pye, J. P. 1979, \apjl, 234, L55 

\bibitem[Stein \& Schwartz(1972)]{ss72}
Stein, R. F. \& Schwartz, R. A. 1972, \apj, 177, 807 

\bibitem[Sturrock (1999)]{str99}
Sturrock, P. A.  1999, \apj, 521, 451 

\bibitem[Suzuki(2002)]{suz02}
Suzuki, T. K. 2002, \apj, 578, 598

\bibitem[Suzuki(2004)]{suz04}
Suzuki, T. K. 2004, \mnras, 349, 1227

\bibitem[Suzuki \& Inutsuka (2005)]{si05}
Suzuki, T. K. \& Inutsuka, S 2005, submitted to ApJL (astro-ph/0506639)

\bibitem[Waldron \& Cassinelli (2000)]{wc00}
Waldron, W. L. \&  Cassinelli, J. P. 2000, \apjl, 548, L45

\bibitem[Weber \& Davis (1967)]{wd67}
Weber, E. J. \& Davis, L. 1967, \apj, 148, 217

\bibitem[Wood et al.(2002)]{wod02}
Wood, B. E., M\"{u}ller, H.-R., Zank, G. P., \& Linsky, J. L. 2002, \apj, 
574, 412

\bibitem[Wood et al.(2005)]{wod05}
Wood, B. E., M\"{u}ller, H.-R., Zank, G. P., Linsky, J. L., \& Redfield 2005, 
ApJL, in press (astro-ph/0506401) 

\bibitem[Yan \& Lazarian (2004)]{yl04}
Yan, H. \& Lazarian, A. 2004, \apj, 614, 757

\end{thebibliography}
\end{document}